# The Amplification in FEL
# with Inhomogeneous Magnetic Field

K.B. Oganesyan

Yerevan Physics Institute, Alikhanyan National Science Lab, Yerevan, Armenia

bsk@yerphi.am

The gain in a plane wiggler with inhomogeneous magnetic field is calculated.. It is shown, that the account of inhomogenity of the magnetic field leads to appearance of additional peaks in the amplification.

## 1. INTRODUCTION

The free electron lasers (FEL) are high power tunable sources of coherent radiation that are used in science research, for heating of plasma, in the physics of condensed media, in atomic, molecular and optical physics, in biophysics, biochemistry, biomedical engineering etc. The radiation produced by present-day FELs has a range from the millimeter to the X-rays waves, which no other similar high intensity tunable sources cover [1, 2]. In modern science this branch is of interest both from the viewpoint of fundamental research and application.

In FELs [3, 4] the kinetic energy of relativistic electrons moving through the spatially modulated magnetic field of a wiggler is used for production of coherent radiation. The frequency of radiation is determined by the energy of electrons, the space period and the magnetic field strength of a wiggler. This permits a retuning of FEL to be made in a wide range as opposed to the atomic and molecular lasers. In a conventional FEL the magnetic field of wiggler is constant, but it is inhomogeneous in the transverse direction [5]. It is important to take into account this inhomogeneity that causes a complicated motion of electrons: fast undulator oscillations along the wiggler axis and slow strophotron ones [6–14] in the transverse direction. Recently [15] the spectral distribution of spontaneous emission in plane wiggler with inhomogeneous magnetic field was calculated.

In the present paper we describe the amplification in FEL with inhomogeneous magnetic field.

## 2. SPONTANEOUS RADIATION

The equations of motion in inhomogeneous magnetic field were described in recent paper [15] with vector potential [16]

$$\mathbf{A}_w = -\frac{H_0}{q_0} \cosh q_0 x \sin q_0 z \cdot \mathbf{j}, \tag{1}$$

and magnetic field

$$H_x = H_0 \cosh q_0 x \cos q_0 z, \quad H_y = 0, \quad H_z = -H_0 \sinh q_0 x \sin q_0 z. \tag{2}$$

The spectral distribution of spontaneous emission has a form

$$\frac{d\varepsilon}{d\omega do} = \frac{e^2 \omega^2 \Omega^2 T^2}{8\pi^2 q_0^2} \sum_{n,m,k=-\infty}^{\infty} \left(I_{n+1,k,m} - I_{n,k,m}\right)^2 \left(\frac{\sin u}{u}\right)^2, \tag{3}$$

where

$$\begin{bmatrix} u = \frac{T}{2}\left[\omega\left(\frac{1}{2\gamma^2} + \frac{\Omega^2}{2q_0^2}\right) - (2n+1)q_0 - 2m\Omega\right], \\ I_{n,k,m} = J_{n-k}(Z_1) J_{\frac{k+m}{2}}(Z_2) J_{\frac{k-m}{2}}(Z_2), \\ Z_1 = \frac{\omega\Omega^2}{4q_0^3}, Z_2 = \frac{\omega a_0^2 \Omega^2}{4q_0}. \end{bmatrix} \tag{4}$$

$H_0$ is the strength of magnetic field, $q_0 = 2\pi/\lambda_0$, $\lambda_0$ – undulator period, $\mathbf{j}$ – unit vector in $y$ direction. Equation (3) describe radiation spector consisted of superposition of lines located on combined frequencies of odd harmonics $(2n+1)\omega_{рез,онд}$ of undulator resonance frequency and even harmonics $2m\omega_{рез,стр}$ of strophotron resonance frequency, where $m, n = 0, 1, 2, 3, \cdots$. and

$$\omega_{рез,онд} = \frac{2\gamma^2 q_0}{1+\gamma^2 \frac{\Omega^2}{q_0^2}}, \quad \omega_{рез,стр} = \frac{2\gamma^2 \Omega}{1+\gamma^2 \frac{\Omega^2}{q_0^2}}. \tag{5}$$

Here we take into account the following notifications $1-v = \frac{1}{2\gamma^2}$, where $\gamma = \frac{\varepsilon}{mc^2}$ is the relativistic factor, $c$ is speed of light, $\varepsilon$ is electron energy, $\Omega = \frac{eH_0}{\sqrt{2\varepsilon}}$ is the frequency and $a_0 = \sqrt{x_0^2 + \frac{\alpha^2}{\Omega^2}}$ the amplitude of strophotron oscillations.

The above results are obtained in approximation

$$a_0 q_0 < 1, \quad \frac{\Omega}{q_0} < 1, \quad a_0 \Omega < 1. \tag{6}$$

## 3. GAIN

The gain may be found from the expression for spontaneous radiation (3) with the help of the Madey theorem [17]. To avoid the use of some assumptions made at obtaining of these general relations for the spontaneous and stimulated radiations, we preferred to obtain the gain immediately from the equations of motion.

Let an electromagnetic wave with vector potential propagates along $z$ axis (wiggler axis)

$$\mathbf{A}_w = -\frac{E_0}{\omega} \sin \omega (t - z) \mathbf{i}, \tag{7}$$

where $\omega$ is the frequency of electromagnetic wave, $E_0$ the strength of electric field, $\mathbf{i}$ an unit vector in the $x$ direction.

The equations of electron motion in the wiggler (2) and electromagnetic wave (7) fields are

$$\begin{cases} \dfrac{dp_x}{dt} = -eH_0 q_0 x v_y \sin q_0 z + eE_0 (1 - v_z) \cos \omega (t - z), \\ \dfrac{dp_y}{dt} = eH_0 \left[ v_z \left( 1 + \dfrac{q_0^2 x^2}{2} \right) \cos q_0 z + q_0 v_x x \sin q_0 z \right], \\ \dfrac{dp_z}{dt} = -eH_0 v_y \left( 1 + \dfrac{q_0^2 x^2}{2} \right) \cos q_0 z + eE_0 v_x \cos \omega (t - z), \end{cases} \tag{8}$$

and the energy change is

$$\frac{d\varepsilon}{dt} = e\mathbf{v}\mathbf{E} = eE_0 v_x \cos \omega (t - z). \tag{9}$$

The linear gain (field independent) is determined by the second order electric field strength $(\propto E_0^2)$ and is obtained from equation (9):

$$\frac{d\varepsilon}{dt} = eE_0 v_x^{(1)} \cos \omega \left( t - z^{(0)} \right) + eE_0 \omega v_x^{(0)} z^{(1)} \sin \omega \left( t - z^{(0)} \right). \tag{10}$$

To obtain the gain one needs to find the first-order corrections in the field $(\propto E_0)$ $x^{(1)}(t)$, $z^{(1)}(t)$ to $x^{(0)}(t)$ (11) and $z^{(0)}(t)$ (15) [15].

The first-order corrections obtained from (8) satisfy the following equations:

$$\begin{cases} \dfrac{dp_x^{(1)}}{dt} = -\varepsilon_0 \Omega^2 x^{(1)} + eE_0\left(1 - v_z^{(0)}\right)\cos\omega\left(t - z^{(0)}\right), \\ \dfrac{dp_z^{(1)}}{dt} = eE_0 v_x^{(0)} \cos\omega\left(t - z^{(0)}\right). \end{cases} \qquad (11)$$

Then we find $x^{(1)}(t)$, $z^{(1)}(t)$ from equations (11 26) and using the expressions for $x^{(0)}(t)$ (11) and $z^{(0)}(t)$ (15) [15] obtain an expression for the energy radiated by the electron $\Delta\varepsilon = \int_0^T \dfrac{d\varepsilon}{dt} dt$ during $T$ time of flight through the undulator, as well as for the gain

$$G = \dfrac{4\pi N_e}{E_0^2} \Delta\varepsilon, \qquad (12)$$

where $N_e$ is the concentration of electrons in the beam.

Since all these calculations are simple, but laborious and cumbersome, here we give only the resultant expression:

$$G = \dfrac{e^2 \omega^2 \Omega^2 N_e T^3}{4\pi q_0^2 \gamma^2}\left(1 + \gamma^2 \dfrac{\Omega^2}{q_0^2}\right) \sum_{n,m,k}^{\infty} \left(I_{n+1,m,k} - I_{n,m,k}\right)^2 \dfrac{d}{du}\left(\dfrac{\sin u}{u}\right)^2, \qquad (13)$$

where the notations (4) were used. Equation (13) describes the gain comprising the superposition of spectral lines, localized at combinated frequencies of odd harmonics of $(2n+1)\omega_{\text{res,und}}$ undulator resonant frequency and even harmonics of $2m\omega_{\text{res,str}}$ strophotronic resonant frequency, where $m, n = 0, 1, 2, 3, \ldots$, and $\omega_{\text{res,str}}$ and $\omega_{\text{res,und}}$ are determined by formula (5).

## 4. CONCLUSION

It is shown that as a result of the allowance for magnetic field inhomogeneity, some additional peaks appear in the gain. The peaks of the gain are localized at combinated frequencies of the odd harmonics of $(2n+1)\omega_{\text{res,und}}$ undulator resonant frequency and even harmonics of $2m\omega_{\text{res,str}}$ strophotronic resonance frequency. In case of an undulator with constant magnetic field the peaks are localized at odd harmonics of undulator resonant frequency $\omega_{\text{res,und}} = \dfrac{2\gamma^2 q_0}{1 + \gamma^2 \Omega^2/q_0^2}$, and in case of a strophotron they are localized at odd harmonics of strophotronic resonant frequency $\omega_{\text{res,str}} = \dfrac{2\gamma^2 \Omega}{1 + \gamma^2 \Omega^2/q_0^2}$. One may conclude thus, that due to the presence of inhomogeneity in the magnetic field in the plane wiggler these two systems (wiggler with the constant magnetic field and the strophotron) are integrated in one unit and there appear peaks in the gain at combined (odd undulator and even strophotronic) resonant frequencies.